\documentclass[letterpaper,twocolumn]{revtex4}
\usepackage[latin9]{inputenc}
\usepackage{graphicx}
\usepackage{esint}

\makeatletter

\@ifundefined{textcolor}{}
{%
 \definecolor{BLACK}{gray}{0}
 \definecolor{WHITE}{gray}{1}
 \definecolor{RED}{rgb}{1,0,0}
 \definecolor{GREEN}{rgb}{0,1,0}
 \definecolor{BLUE}{rgb}{0,0,1}
 \definecolor{CYAN}{cmyk}{1,0,0,0}
 \definecolor{MAGENTA}{cmyk}{0,1,0,0}
 \definecolor{YELLOW}{cmyk}{0,0,1,0}
}

\makeatother

\begin{document}

\title{Sympathetic cooling in a large ion crystal}

\author{Guin-Dar Lin$^{1}$ and Luming Duan$^{2,3}$}

\affiliation{$^{1}$Center for Quantum Science and Engineering, and Department
of Physics, National Taiwan University, Taipei 10617, Taiwan\\
$^{2}$Department of Physics, University of Michigan, Ann Arbor,
Michigan 48109, USA\\
 $^{3}$Center for Quantum Information, IIIS, Tsinghua University,
Beijing 100084, China}
\begin{abstract}
We analyze the dynamics and steady state of a linear ion array when
some of the ions are continuously laser cooled. We calculate the ions'
local temperature measured by its position fluctuation under various
trapping and cooling configurations, taking into account background
heating due to the noisy environment. For a large system, we demonstrate
that by arranging the cooling ions evenly in the array, one can suppress
the overall heating considerably. We also investigate the effect of
different cooling rates and find that the optimal cooling efficiency
is achieved by an intermediate cooling rate. We discuss the relaxation
time for the ions to approach the steady state, and show that with
periodic arrangement of the cooling ions, the cooling efficiency does
not scale down with the system size.
\end{abstract}
\maketitle

\section{Introduction\label{sec:Introduction}}

Trapped ions constitute one of the leading systems for implementation
of quantum computation. Numerous advances have been achieved in this
system, including realization of faithful quantum gates \cite{Sorensen_PRL1999,Milburn_FP2000,Sorensen_PRA2000,Leibfried_Nature2003,Myerson2008,Burrell2010,Harty2014},
preparation of many-body quantum states \cite{Sackett_Nature2000,Leibfried_Nature2005,Haffner_Nature2005,Monz2011,Noguchi2012,Jurcevic2014,Lanyon2014,Northup2015},
and quantum teleportation \cite{Riebe_Nature2004,Olmschenk_Science2009}.
There are also developments to scale up this system, based on either
ion shuttling \cite{Kielpinski2002,Bowler2012,Walther2012} or quantum
networks \cite{Duan_PRA2006,Moehring_Nature2007,Duan2010,Northup2014,Monroe2014,Hucul2015}).

In a typical ion trap, the ions are first Doppler cooled and form
a crystal. Most of the quantum computation experiments use a one-dimensional
ion crystal. The ions may be subjected to further sub-Doppler cooling,
such as sideband cooling. However, the difficulty of sideband cooling
scales up with the number of phonon modes, which increase with the
number of ions \cite{Wineland_JNIST1998,Leibfried_RMP2003,Lin_EPL2009}.
It has been shown that in principle high-fidelity quantum computation
can be achieved even at the Doppler temperature by employing the ions'
transverse phonon modes \cite{Zhu_PRL2006,Lin_EPL2009}. In a real
experimental setup, the ions are subject to substantial background
heating. For long-time quantum computation, to have the ions constantly
remain at a certain temperature, it requires sympathetic cooling \cite{Larson1986,Kielpinski2000},
in which case a subset of ions (cooling ions) are continuously laser
cooled, bringing down the temperature of other ions (the computational
ions) through the heat propagation enabled by the Coulomb interaction
in the ion crystal. Sympathetic cooling has been studied for small
systems with a few ions \cite{Barrett2003,Home2009,Brown2011}.

In this paper, we study the effectiveness of sympathetic cooling in
a large one-dimensional ion crystal. Although in general temperature
is not well defined for this system as it does not reach a thermal
equilibrium state, as a relevant indicator for quantum computation,
we measure the local ``temperature'' of the ions through their average
position fluctuation (PF) $\delta x_{i}^{\xi}\equiv\sqrt{\langle x_{i}^{\xi2}\rangle}$
(for the $i$th ion) with $\xi=x,y$ for the transverse phonon modes
and $\xi=z$ for the the axial modes. This position thermal fluctuation
is an important indicator for fidelity of quantum gates. We discuss
two different arrangements of the cooling ions: edge cooling and periodic-node
cooling. In the former case, the ions at the two edges of an ion array
are continuously laser cooled. In the latter case, the cooling ions
are distributed evenly and periodically in the ion chain. We show
that the periodic-node cooling is much more effective than the edge
cooling. For a large crystal, the edge cooling becomes very inefficient.
We then discuss the nontrivial dependence of the local temperature
of the computational ions on the cooling rate of the cooling ions.
A large cooling rate does not necessarily lead to more efficient cooling
of the computational ions. Instead, there is an intermediate optical
cooling rate, in agreement with our previous observation \cite{Lin_NJP2011}.
We finally investigate the time scale for the system to reach the
steady state, which in general differs from the thermal equilibrium
state \cite{Lin_NJP2011}.

This paper is organized as follows. In Sec. \ref{sec:Formalism},
we present the Heisenberg-Langevin equations to describe the driven
dynamics of a many-ion array and provide their formal exact solutions.
In Sec. \ref{sec:Steady-state-distribution}, we discuss the motional
steady states of the ions under background heating and continuous
sympathetic cooling on the cooling ions. In Sec. \ref{sec:Architecture},
we study different cooling configurations and discuss the corresponding
cooling efficiency. In Sec. \ref{sec:Cooling-rate} we investigate
how the cooling performance of the sympathetic cooling depends on
the laser cooling rate. In Sec. \ref{sec:Time-evolution}, we study
the relaxation dynamics of the cooling process, and discuss the time
scale of relaxation as well as its scaling with the system size. Finally,
we summarize the major findings in Sec. \ref{sec:Conclusion}.

\section{Formalism \label{sec:Formalism}}

Consider an ion string confined in an RF trap with an effective static
potential $\mathcal{V}(\mathbf{r})=\frac{1}{2}m\omega_{x}^{2}(x^{2}+y^{2})+V(z)$.
For a small crystal, the axial confinement is usually approximated
by $V(z)=\frac{1}{2}m\omega_{z}^{2}z^{2}$ with $\omega_{z}\ll\omega_{x}$
so that the one dimensional alignment is stabilized. For a large crystal,
the axial potential might take an anharmonic form \cite{Lin_EPL2009}.
Trapped ions have collective motion around their classical equilibrium
positions. Assuming that each ion is coupled to its respective thermal
bath (corresponding to either cooling or background heating), we describe
the driven ion array by the following Heisenberg-Langevin equations:
\begin{equation}
\left\{ \begin{array}{ccl}
\dot{x}_{i}^{\xi} & = & p_{i}^{\xi}\\
\dot{p}_{i}^{\xi} & = & -\sum_{j}A_{ij}^{\xi}x_{j}^{\xi}-\gamma_{i}^{\xi}p_{i}^{\xi}+\sqrt{2\gamma_{i}^{\xi}}\zeta_{i}^{\xi}(t)
\end{array}\right.,\label{eq:Langevin}
\end{equation}
where $i,j=1,...,N$ are ion indices, $\xi=x,y$,$z$ stands for the
mode directions, and $A_{ii}^{\xi}=\beta_{i}^{\xi}-\sum_{j(\neq i)}\frac{C_{\xi}}{\left\vert z_{j}^{0}-z_{i}^{0}\right\vert ^{3}},$
$A_{ij}^{\xi}=\frac{C_{\xi}}{\left\vert z_{j}^{0}-z_{i}^{0}\right\vert ^{3}}(i\neq j)$
with $\beta_{i}^{x,y}=\omega_{x}^{2}$, $\beta_{i}^{z}=\partial^{2}V/\partial z_{i}^{2}$,
$C_{x,y}=1$, $C_{z}=-2$, and $z_{i}^{0}$ denotes the $i$th ion's
axial equilibrium position. We take the ion spacing $d_{0}$ as the
length unit\footnote{Here the choice of $d_{0}$ is somewhat arbitrary as long as it characterizes
the length scale of the inter-ion spacing. In this article we define
$d_{0}$ differently in various situations. For instance, in a small
harmonic trap ($N=20$), we we choose $d_{0}$ to be the smallest
spacing in the middle of the chain. In a large nonuniform ion crystal
($N=121$), we choose $d_{0}=\frac{1}{100}\sum_{i=11}^{110}(z_{i+1}^{0}-z_{i}^{0})/100$,
a mean value of all ion spacings except that 10 large ones on the
edges are excluded.}, $e^{2}/d_{0}$ as the energy unit, and $\omega_{0}\equiv\sqrt{e^{2}/(md_{0}^{3})}$
as the frequency unit so that the quantities in Eq. (\ref{eq:Langevin})
is dimensionless. For the convenience of discussion, we drop the superscript
$\xi$. Since the transverse and axial modes are decoupled, the derivation
simply applies to any direction. A random kick $\zeta_{i}(t)$ associated
with the driving rate $\gamma_{i}$ can be expressed as $\zeta_{i}=-i\sum_{k}\sqrt{\frac{\omega_{k}}{2}}G_{ik}(b_{k}-b_{k}^{\dagger})$
(in units of $\sqrt{\hbar m\omega_{0}}$), where $G$ is the canonical
transformation matrix which diagonalizes $A$, i.e. $G^{\top}AG=A^{D}$
is diagonalized, and $b_{k}$ is the bosonic field operator of the
$k$th motional mode with frequency $\omega_{k}$. For a Markovian
bath, $b_{k}(t)$ satisfies $\bigl\langle b_{k}^{\dagger}(t_{1})b_{k^{\prime}}(t_{2})\bigr\rangle=n_{k}^{B}(T)\delta_{kk^{\prime}}\delta(t_{1}-t_{2})$
with $n_{k}^{B}(T)\equiv\left[\exp(\omega_{k}/T)-1\right]^{-1}$ the
phonon number of the $k$th mode for a given temperature $T$ (in
units of $\hbar\omega_{0}/k_{B}$). It is then straightforward to
show that the correlation of the driving force is given by $\bigl\langle\zeta_{i}(t)\zeta_{j}(t^{\prime})\bigr\rangle=\delta_{ij}\delta(t-t^{\prime})\sum_{k}\omega_{k}G_{ik}^{2}\bigl(n_{k}^{B}(T)+\frac{1}{2}\bigr)$.
In our current case, where each ion couples to an independent reservoir
$T_{i}$, it is reasonable to assume that the ion $i$ feels a local
bath with $\bigl\langle\zeta_{i}(t)\zeta_{j}(t^{\prime})\bigr\rangle=\delta_{ij}\delta(t-t^{\prime})\Theta_{i}(T_{i})$
and $\Theta_{i}(T_{i})\equiv\sum_{k}\omega_{k}G_{ik}^{2}\bigl(n_{k}^{B}(T_{i})+\frac{1}{2}\bigr)$.
The solution to Eq. (\ref{eq:Langevin}) is given by $\mathbf{q}(t)=e^{-\Omega t}\mathbf{q}(0)+\int_{0}^{t}d\tau e^{\Omega(\tau-t)}\mathbf{\eta}(\tau)$,
where $\mathbf{q}\equiv(x_{1},x_{2},...;p_{1},p_{2},...)^{\top}=\left[\begin{array}{c}
\left\{ x_{i}\right\} \\
\left\{ p_{i}\right\} 
\end{array}\right]$, $\eta(t)\equiv\left[\begin{array}{c}
\left\{ 0\right\} \\
\left\{ \sqrt{2\gamma_{i}}\zeta_{i}\right\} 
\end{array}\right]$, and $\Omega\equiv\left[{\normalcolor \begin{array}{cc}
{\normalcolor 0} & {\normalcolor -I}\\
{\normalcolor \bigl[A_{ij}\bigr]} & \bigl[\gamma_{i}\delta_{ij}\bigr]
\end{array}}\right]$ is a $2N\times2N$ matrix which can be diagonalized as $\bigl[U^{-1}\Omega U\bigr]_{\alpha\beta}=\lambda_{\alpha}\delta_{\alpha\beta}$.
We then obtain the variation of operators $x_{i}$ and $p_{i}$:
\begin{eqnarray}
\bigl\langle q_{\mu}^{2}\bigr\rangle & = & \sum_{s=1}^{N}\sum_{\alpha,\beta=1}^{2N}U_{\mu\alpha}U_{\mu\beta}\Biggl(e^{-(\lambda_{\alpha}+\lambda_{\beta})t}\biggl[\bigl\langle x_{s}^{2}(0)\bigr\rangle U_{\beta s}^{-1}U_{\alpha s}^{-1}\nonumber \\
 & + & \bigl\langle p_{s}^{2}(0)\bigr\rangle U_{\beta,s+N}^{-1}U_{\alpha,s+N}^{-1}\biggr]\label{eq:Variance}\\
 & + & \Bigl(1-e^{-(\lambda_{\alpha}+\lambda_{\beta})t}\Bigr)\frac{2\gamma_{s}\Theta_{s}}{\lambda_{\alpha}+\lambda_{\beta}}U_{\beta,s+N}^{-1}U_{\alpha,s+N}^{-1}\Biggr),\nonumber 
\end{eqnarray}
where $\mu=1,2,...,N$ correspond to $x$-operators, $\mu=N+1,N+2,...,2N$
correspond to $p$-operators.

For trapped ion quantum computing, the computational fidelity is determined
by the ion PF $\delta x_{i}^{\xi}\equiv\sqrt{\bigl\langle x_{i}^{\xi2}\bigr\rangle}$
(denoted by $\delta x_{i}$ and $\delta z_{i}$ for transverse and
axial motion, respectively). When the quantum gate is operated by
means of the transverse modes, the estimated infidelity is $\delta F_{i}^{x}\sim\pi^{2}\eta_{i}^{4}/4$
\cite{Sorensen_PRA2000,Zhu_PRL2006,Lin_EPL2009}, where the Lamb-Dicke
parameter $\eta_{i}\sim\left\vert \Delta\mathbf{k}\right\vert \delta x_{i}$
with $\Delta\mathbf{k}\parallel\hat{x}$ the wavevector difference
of the two Raman beams. Another possible source of error comes from
the spatial non-uniformity of the laser intensity when a single beam
addresses a specific ion; the ion's axial motion results in variation
of the actual Rabi frequency. This error is estimated by $\delta F_{i}^{z}\sim\pi^{2}(\delta z_{i}/w)^{4}/2$
given that the laser beam's Rabi frequency is approximated by a Gaussian
profile $\Omega(z)\propto e^{-((z-z_{i}^{0})/w)^{2}}$ with width
$w$ \cite{Lin_EPL2009}. Both of the gate errors are determined by
the position thermal fluctuation $\delta x_{i}$ or $\delta z_{i}$
of the ions. So, in the following discussion we focus on the distribution
of the ion position fluctuation $\delta x_{i}$ or $\delta z_{i}$
in the array.

\section{Steady-state distribution\label{sec:Steady-state-distribution}}

\subsection{Thermal equilibrium\label{sub:Thermal-equilibrium}}

\begin{figure}
\begin{centering}
\includegraphics[width=8.5cm]{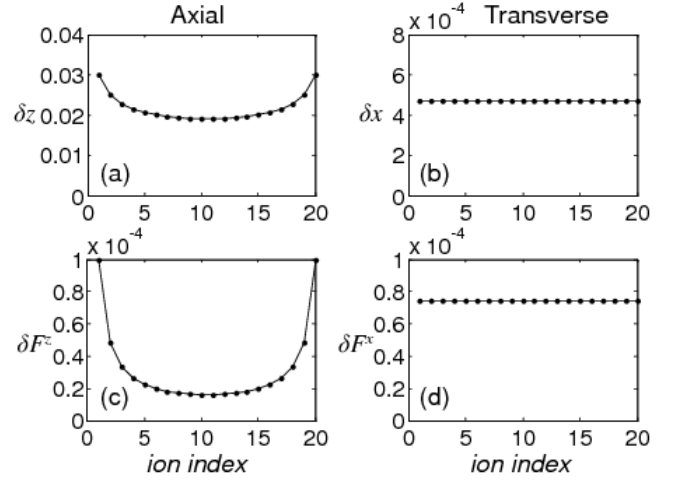} 
\par\end{centering}

\protect\caption{(a)(b) The thermal-equilibrium distributions of the averaged ion PF
in units of $d_{0}$ and (c)(d) the associated computational infidelity
corresponding to the axial and transverse motion at the Doppler temperature
$k_{B}T_{D}/\hbar=2\pi\times9.9$MHz for 20 $^{171}$Yb$^{+}$ ions.
Other parameters: $\omega_{x}=2\pi\times5.1$MHz, $\omega_{z}=2\pi\times34$kHz,
$|\Delta\mathbf{k}|d_{0}=157$ with $d_{0}=10\mu$m, the minimal ion
spacing in the middle of the chain.}
\label{fig:thermal_equilibrium} 
\end{figure}

We first look at the thermal-equilibrium distribution of the ion chain
when the whole system is driven by a thermal field with a well-defined
temperature. From $x_{i}=\alpha\sum_{k}G_{ik}\sqrt{\frac{1}{2\omega_{k}}}(a_{k}^{\dagger}+a_{k})$
where $\alpha\equiv\sqrt{\hbar/(m\omega_{0})}/d_{0}$ is the length
conversion factor and $a_{k}$ ($a_{k}^{\dagger}$) is the annihilation
(creation) operator of mode $k$, we obtain $\bigl\langle x_{i}^{2}\bigr\rangle=\alpha^{2}\sum_{k}\frac{G_{ik}^{2}}{\omega_{k}}\bigl(n_{k}^{B}(T)+\frac{1}{2}\bigr)=\alpha^{2}\sum_{k}\frac{G_{ik}^{2}}{2\omega_{k}}\coth\Bigl(\frac{\omega_{k}}{2T}\Bigr)$.
In Fig. \ref{fig:thermal_equilibrium} we show the distribution of
$\delta x_{i}$ and $\delta z_{i}$ in a harmonic trap for both the
axial and transverse motion at the Doppler temperature $T_{D}$, and
their contribution to the corresponding gate infidelities \cite{Lin_EPL2009}.
In this case, the axial fluctuation $\delta z_{i}$ varies in space,
suggesting that the longitudinal motion of the whole ion chain is
``more collective'' and relies on the global geometry. Supposing
that a single ion is subjected to a different temperature, its longitudinal
movement does not directly reveal information of the temperature associated
with the local bath because neighboring ions subjected to their own
baths may interfere through collective modes. On the contrary, its
transverse movement directly reflects the local temperature. This
is because the axial and transverse modes are decoupled, and for each
ion the energy scale set by the transverse confinement $\hbar\omega_{x}$
is dominant over other scales. Note that the diagonal terms of $A^{x}$
are more significant than the off-diagonal ones, meaning that the
``local modes'' defined by $x_{i}$ and $p_{i}^{x}$ can be discussed
separately from those at different sites, with only small corrections
due to inter-ion coupling. This is where the concept of a ``local
temperature'' for a single ion starts to make sense. Such consideration
has also motivated our investigation about the validity of classical
thermal transportation for the trapped ion system \cite{Lin_NJP2011}.
Each ion can then be approximated as an harmonic oscillator weakly
coupled to others, whose ``local'' phonon occupation number is given
by $n_{i}=\frac{\alpha^{-2}}{2}(\omega_{x}\langle x_{i}^{2}\rangle+\omega_{x}^{-1}\langle p_{i}^{x2}\rangle-1)\approx\alpha^{-2}\omega_{x}\langle x_{i}^{2}\rangle-\frac{1}{2}$.
In the case shown in Fig. \ref{fig:thermal_equilibrium}, PF$=10^{-3}d_{0}$
corresponds to $n_{i}=8.5$ with $\alpha=2.0\times10^{-3}$.\footnote{Throughout this article we choose ytterbium 171 ions spaced by $d_{0}=10\mu$m
as examples, so $\omega_{0}=9.0$MHz and $\alpha=2.0\times10^{-3}$.}

\subsection{Steady-state profile under sympathetic cooling \label{sub:steady-state}}

If different parts of the system make contact with reservoirs at different
temperatures, as relevant for sympathetic cooling, the local temperature
of the ions in the steady state will in general have a non-uniform
spatial profile. In this section, we investigate this steady-state
profile.

We first examine an example where the two edge portions of the ion
chain are continuously laser cooled (we assume Doppler cooling, although
the formalism also applies to other kinds of sympathetic cooling).
The rest of the ion chain is driven by a hot bath corresponding to
the background heating. According to Eq. (\ref{eq:Variance}), in
the long-time limit a steady state should be reached, providing a
time-invariant profile of the position fluctuation $\delta x_{i}$
or $\delta z_{i}$ over all the ions.

\begin{figure}
\begin{centering}
\includegraphics[width=8.5cm]{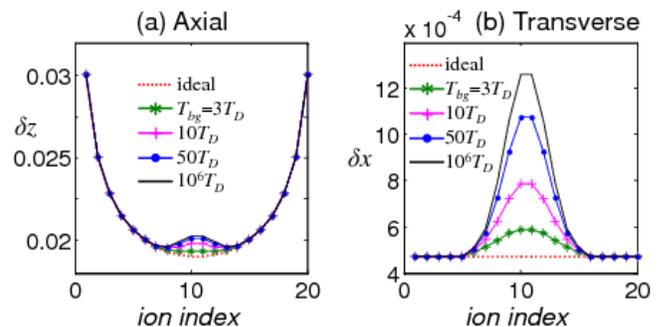} 
\par\end{centering}

\protect\caption{(Color online) The steady-state distributions of the ion PF in a harmonic
trap under different background temperature for a given constant heating
rate. As a comparison, the ideal (no background heating, $\gamma_{bg}=0$)
curves are shown in red dotted lines. }
\label{fig:diff_Tbg} 
\end{figure}

To model the effect of background heating, we assume a small value
for the background driving rate $\gamma_{bg}$ with respect to the
associated environment temperature $T_{bg}$. The value of $T_{bg}$
is hard to quantify; the actual experimentally accessible parameter
is the creation rate of phonons for a given motional mode $k$, that
is, $\gamma_{bg}n_{k}^{B}\sim\gamma_{bg}T_{bg}/\omega_{k}$. To simplify
our discussion, we treat the generated phonon numbers approximately
the same around the range of all transverse (axial) modes. In other
words, the background heating is now only characterized by $\kappa\equiv\gamma_{bg}T_{bg}$.
Nevertheless, for a given value of $\kappa$ we still have the freedom
to vary $T_{bg}$ (and hence $\gamma_{bg}$) while keeping $\kappa$
a constant parameter. As an example, we here consider an $N=20$ chain
with $5$ ions on both ends as cooling ancillas. By denoting the set
of the cooling ancillary ions by $C$ and rest of the chain by $H$,
we take $T_{i}=T_{bg}$, $\gamma_{i}=\kappa/T_{bg}$ for $i\in H$
and $T_{i}=T_{D}$, $\gamma_{i}=0.1$ for $i\in C$. We then compare
the resultant steady-state profile of $\delta x_{i}$ and $\delta z_{i}$
under various $T_{bg}$ in Fig. \ref{fig:diff_Tbg} with constant
$\kappa=10^{-4}$, which amounts to a heating rate of about $60$
phonons per second per ion for the lowest axial mode of $2\pi\times34$kHz.
Note that in a real ion trap, a typical heating rate is about $100\sim1000$
photons per second. As expected, the PF of the ancillary ions coincides
with their supposed thermal-equilibrium values at the Doppler temperature
$T_{D}$ while $\delta x_{i}$ and $\delta z_{i}$ show a hump in
the middle part of the distribution due to the background heating.
For $T_{bg}$ set to larger values, the hump grows but asymptotically
converges to a fixed profile, providing an \textit{upper bound} of
the profile. This corresponds to the ``worst\textquotedblright{}
case with the largest contribution to the gate infidelity. In the
following, we only show such upper bounds for all the circumstances
and investigate the discrepancy between these bounds and the fluctuation
profile at the Doppler temperature (corresponding to the perfectly
cooled case).

\section{Comparison of different cooling configurations\label{sec:Architecture}}

In this section, we compare the efficiency of sympathetic cooling
under two cooling configurations: edge cooling and periodic-node cooling.
For each case, we show the results under both harmonic and anharmonic
axial traps. For a large ion crystal, the inhomogeneous ion spacing
under a harmonic trap complicates the gate design and reduces its
fidelity for quantum computation. To overcome this problem, as suggested
in \cite{Lin_EPL2009}, it is better to use anharmonic traps which
can give equal or almost equal spacing for the ions in the chain.
We consider two kinds of anharmonic trap: the one (called the uniform
trap for simplicity of terminology) which gives perfect uniform spacing
for the ions and the quartic trap with potential $V(z)=\frac{1}{2}\alpha_{2}z^{2}+\frac{1}{4}\alpha_{4}z^{4}$
which gives approximate uniform ion spacings. The parameters $\alpha_{2}$
and $\alpha_{4}$ in $V(z)$ are chosen to minimize the variation
of the distribution of the ion spacings in the chain \cite{Lin_EPL2009}.

\subsection{Edge cooling\label{sub:Edge-cooling}}

\begin{figure}
\begin{centering}
\includegraphics[width=8.5cm]{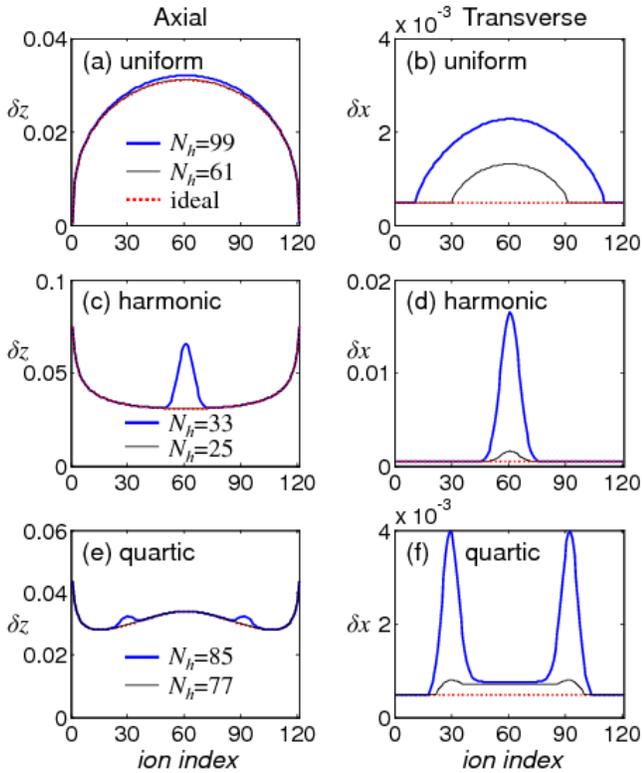} 
\par\end{centering}

\protect\caption{(Color online) The distributions of the axial and transverse PF (in
units of $d_{0}$) for various trap geometries under the edge cooling:
(a)(b) a uniform array, (c)(d) a harmonic trap, (e)(f) a quartic trap.
$(N-N_{h})/2$ ions on each end of the ion chain are Doppler cooled
with a driven rate $\gamma=0.1$. For the uniform array, the spacing
is $d_{0}=10\mu$m; for the harmonic case, $\omega_{z}=2\pi\times8.4$kHz
and $\omega_{x}=2\pi\times5.1$MHz, and for the quartic case $\omega_{2}\equiv\sqrt{|\alpha_{2}|/m}=2\pi\times5$kHz
and $|\alpha_{2}/e^{2}|^{2/3}(\alpha_{2}/\alpha_{4})=-6.2$ such that
$\sum_{i=11}^{110}(z_{i+1}^{0}-z_{i}^{0})/100=d_{0}$. The background
heating rate $\kappa=10^{-4}$ amounts to, for instance, generating
$240$ phonons per second for the lowest harmonic mode $\omega_{z}$.
Other parameters are the same as used in Fig. \protect\ref{fig:thermal_equilibrium}.}
\label{fig:edgecool} 
\end{figure}

First, we show the result with the edge segments of the ions are Doppler
cooled. Fig. \ref{fig:edgecool} shows the final distributions of
$\delta x_{i}$ and $\delta z_{i}$ under three different traps. As
a comparison, the corresponding thermal-equilibrium profiles at $T=T_{D}$
are shown as red dotted curves. To consider how many ions can be cooled
effectively through sympathetic cooling, we show the curves under
different number $N_{h}\equiv n(H)$ of the computational ions which
are subject to the background heating. The axial distribution is shown
in Fig. \ref{fig:edgecool}(a), (c) and (e). In the uniform case,
the curves almost coincide with the ideal thermal equilibrium one
under the Doppler temperature, indicating that the system is almost
perfectly cooled by sympathetic cooling. In our example with the system
size $N=121$, the edge cooling for a uniform ion chain can afford
$N_{h}$ up to $100$ ions, with the maximal $\delta z_{i}$ (occurring
at the middle ion with $i=61$) increased by about $4\%$ compared
with the ideal case. In the harmonic trap, the affordable $N_{h}$
is significantly reduced; the ion PF $\delta z_{i}$ grows very fast
near the chain center as $N_{h}$ exceeds a certain value $(\sim25)$.
A considerable improvement can be found in the quartic case, which
supports up to $N_{h}\sim85$ ions with negligible discrepancy in
the distribution. With even larger $N_{h}$, humps start to form on
two sides instead of being at the chain center. As for the transverse
motion, as shown in Fig. \ref{fig:edgecool}(b), (d), and (f), the
cooling efficiency is in general more vulnerable than that of the
axial motion. Because the transverse motion is typically more localized,
the ancillary ions have vanishing influences on the ions of increasing
distance. It can be observed that although $\delta x_{i}$ for the
edge ions are fixed by the Doppler temperature, the ions away from
the laser cooled ions soon get large $\delta x_{i}$. Therefore, we
expect that it is inefficient to cool the transverse modes with the
edge cooling.

\begin{figure}
\begin{centering}
\includegraphics[width=8.5cm]{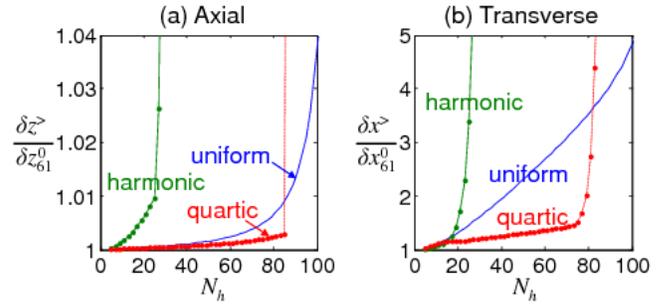} 
\par\end{centering}

\protect\caption{(Color online) The cooling efficiency (in terms of the normalized
PF) as a function of the number of heat-driven ions for (a) the axial
and (b) the transverse motion.}
\label{fig:nedge} 
\end{figure}

The dependence of the cooling efficiency on the number of computational
ions $N_{h}$ is plotted in Fig. \ref{fig:nedge}. To quantify the
cooling efficiency, here we look at the maximal axial (transverse)
position fluctuation $\delta z^{>}$ ($\delta x^{>}$) among all the
ions belonging to $H$ normalized by the middle one's fluctuation
$\delta z_{m}^{0}$ ($\delta x_{m}^{0}$) at the Doppler temperature
$T_{D}$ ($m=61$ for the system size $N=121$). With this definition,
the normalized characteristic fluctuation approaches the unity when
the system reaches the Doppler temperature. With this setup, the sympathetic
cooling works better for the axial modes than the transverse ones
in terms of the gate infidelities $\delta F^{z}$ and $\delta F^{x}$,
which are proportional to $\bigl(\frac{\delta z^{>}}{\delta z_{61}^{0}}\bigr)^{4}$
and $\bigl(\frac{\delta x^{>}}{\delta x_{61}^{0}}\bigr)^{4}$, respectively.
For instance, the infidelity $\delta F^{z}$ is roughly increased
by $16\%$ for $\frac{\delta z^{>}}{\delta z_{61}^{0}}\sim1.04$ but
$\delta F^{x}$ is increased by $16$ times for $\frac{\delta x^{>}}{\delta x_{61}^{0}}\sim2$.
It is interesting to observe that for both the axial and transverse
directions, the curves for the quartic trap rise more slowly than
those for the uniform trap before they suddenly jump up around $N_{h}\sim85$.

\subsection{Periodic-node cooling\label{sub:Periodic-cooling}}

\begin{figure}
\begin{centering}
\includegraphics[width=8.5cm]{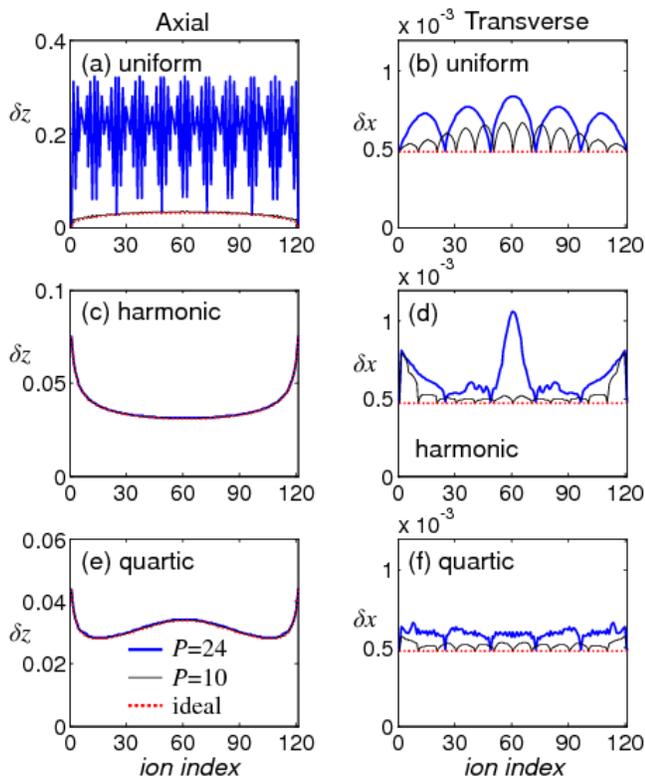} 
\par\end{centering}

\protect\caption{(Color online) The distributions of the axial and transverse PF (in
units of $d_{0}$) under the periodic cooling for (a)(b) a uniform
array, (c)(d) a harmonic trap, (e)(f) a quartic trap. The total number
of ions $N=121$. All parameters are the same as in Fig. \protect\ref{fig:edgecool}
except for the ancilla arrangement. The curve legend for all six panels
is the same and is given in (e).}
\label{fig:periodiccool} 
\end{figure}

As discussed above for the edge cooling, if we impose an efficiency
threshold, there must be a limit on $N_{h}$ beyond which the system
cannot be effectively cooled. For long ion chains, therefore, a different
spatial arrangement of the cooling ions must be considered. Here,
we discuss an improved configuration where the ancillary cooling ions
are distributed periodically and evenly in the ion chain. We investigate
how the period (the number of computational ions between two adjacent
cooling ions/nodes) influences the performance of sympathetic cooling.
We still take the ion number $N=121$ as an example and only Doppler
cool the $1$st, $(1+P)$th, $(1+2P)$th, ..., $N$th ions with a
period $P$ that factorizes $120$. In Fig. \ref{fig:periodiccool}
we show the resultant distribution of $\delta x_{i}$ and $\delta z_{i}$
under three different trapping potentials. Unlike the edge cooling
case, a uniform chain has no good performance under the periodic-node
cooling. As the reason will be revealed later in Sec. \ref{sec:Cooling-rate},
this is because the cooling rate $\gamma=0.1$ is not an optimal choice.
As for the axial motion in the harmonic and quartic cases shown in
Fig. \ref{fig:periodiccool}(c) and (e), the curves are almost identical
to the ideal ones even with a large period $P=24$ (about $5\%$ of
the ions are used for sympathetic cooling in this case). For the transverse
direction shown in Fig. \ref{fig:periodiccool}(b), (d), and (f),
$\delta x_{i}$ is significantly suppressed compared to those under
the edge cooling configuration. Although the detailed distribution
depends on the trapping potential, the maximum $\delta x_{i}$ is
no more than two ($1.25$) times of $\delta x_{i}$ for the ideal
case under a large period $P=24$ ($P=10$).

\begin{figure}
\begin{centering}
\includegraphics[width=8.5cm]{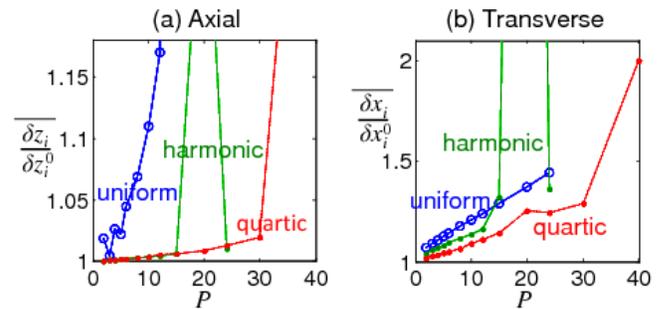} 
\par\end{centering}

\protect\caption{(Color online) The efficiency of the periodic cooling with different
periods of ancilla arrangement for (a) the axial and (b) the transverse
motion. }
\label{fig:nperiod} 
\end{figure}

We plot the cooling efficiency against the period $P$ in Fig. \ref{fig:nperiod}.
Here the efficiency is characterized by $\overline{\frac{\delta z_{i}}{\delta z_{i}^{0}}}\equiv\frac{1}{n(H)}\sum_{i\in H}\frac{\delta z_{i}}{\delta z_{i}^{0}}$
(similarly for $\overline{\frac{\delta x_{i}}{\delta x_{i}^{0}}}$).
Note that in the uniform case, the efficiency becomes worse due to
the improper choice of $\gamma=0.1$. For the axial modes shown in
Fig. \ref{fig:nperiod}(a), the efficiency in the harmonic case is
as good as that in the quartic case except at $P=20$, where the ion
PF suddenly jumps out of the good range in the harmonic case. For
the transverse modes (Fig. \ref{fig:nperiod}(a)), the three trap
potentials do not show dramatic differences for $P<15$, but in general
the quartic curve still shows the slowest increase in the ion PF as
$P$ gets larger. The exception with a sudden jump of the PF at $P=20$
is somewhat related to a particular phonon eigenmode structure for
the harmonic trap. Such a eigenmode happens to have a few nodal points
coincident with the sites of cooling ions. Therefore this mode cannot
be cooled effectively. This can be circumvented by arranging cooling
ions asymmetrically with respect to the trap center. On the other
hand, if some of the ions happen to be of large PF in one normal mode,
cooling these ions effectively cools this mode. So it might be ideal
to choose to cool those ions whose amplitudes are large in most of
the eigenmodes.

\section{Influence of cooling rates\label{sec:Cooling-rate}}

\begin{figure}
\begin{centering}
\includegraphics[width=8.5cm]{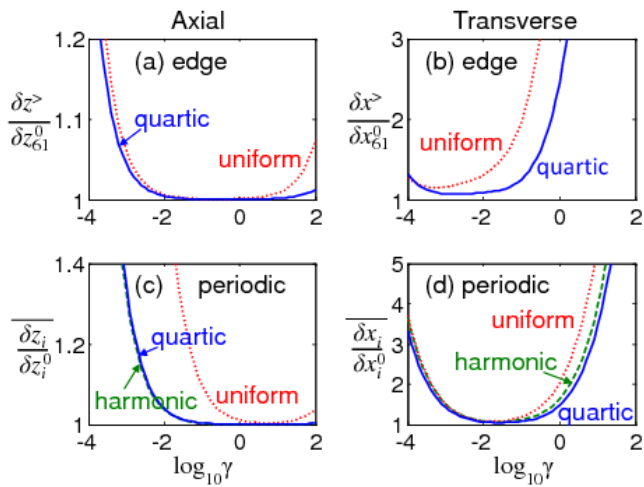} 
\par\end{centering}

\protect\caption{(Color online) The cooling efficiency for (a) the axial modes and
(b) the transverse modes under edge cooling; (c) the axial modes and
(d) the transverse modes under periodic-node cooling. The system size
$N=121$. In (a) and (b), these curves correspond to $N_{h}=41$.
In (c) and (d), the curves correspond to $P=10$. Other parameter
are the same as previously discussed.}
\label{fig:coolrate} 
\end{figure}

In this section, we discuss the significance of the driving rate $\gamma$
of the Doppler cooled ions. Intuitively, we would expect that the
system can be cooled more efficiently when the driving rate $\gamma$
gets larger. Our calculation shows that this is however not the case.
We study the efficiency with varied $\gamma$ under the same background
heating rate $\kappa$. The efficiency characterized by the corresponding
(normalized) position fluctuation is plotted in Fig. \ref{fig:coolrate}
as a function of $\gamma$. Surprisingly, for all the circumstances
we consider, the ion position fluctuation first decreases as the driving
rate rises in the small $\gamma$ regime, approaching to a minimum
when $\gamma$ is moderate, and then increases again when $\gamma$
becomes strong. This suggests that the driving rate has an optimal
window for cooling. The fact that the efficiency does not go better
with strong cooling rates seems counter-intuitive in the first place.
But this finding is consistent with our previous work \cite{Lin_NJP2011}.
The reason is that, when the driving rate is larger than the inverse
of the timescale needed for propagation, the ion is kicked from random
directions so frequently that the effects of succeeding kicks cancel
out before the first kick is about to \textquotedblleft transfer\textquotedblright{}
to its neighbors. If the rate matches the propagation timescale in
order of magnitude, the cooling efficiency gets optimal.

Furthermore, these curves do not reach the minima at the same $\gamma$;
the optimized cooling rate depends on the trapping potentials, cooling
configurations, and which direction of the motion is considered. For
the edge cooling (Fig. \ref{fig:coolrate}(a) and (b)), the most efficient
window of $\gamma$ for cooling axial modes lies in the range from
$0.1$ to $1$, both for the uniform and the quartic potentials. With
the same rate, the (normalized) transverse PF becomes significantly
larger than unity for these two geometries. For the periodic-node
cooling (Fig. \ref{fig:coolrate}(c) and (d)), both the curves for
the harmonic and the quartic potentials are nearly identical, with
the optimal window lies in the range from $\gamma\sim0.1$ to $10$
for the axial direction and from $\gamma\sim0.02$ to $0.05$ for
the transverse direction. By compromising the optimal windows for
both the directions, $\gamma\simeq0.1$ sounds a suitable choice.

\section{Relaxation dynamics to the steady state \label{sec:Time-evolution}}

\begin{figure}
\begin{centering}
\includegraphics[width=8.5cm]{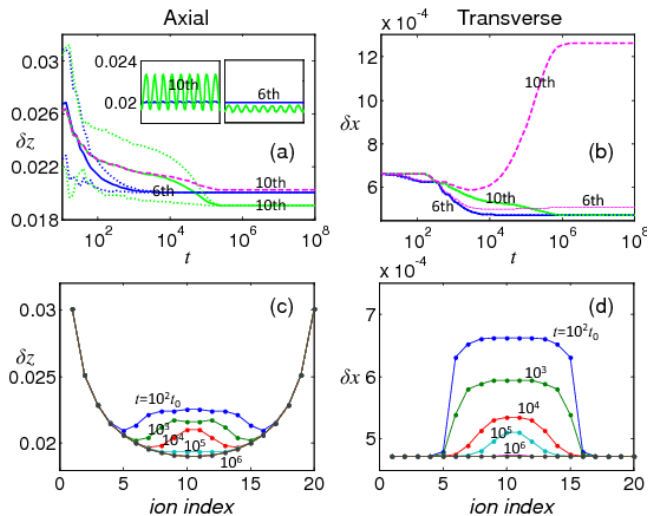} 
\par\end{centering}

\protect\caption{(Color online) (a)(b) The temporal profiles of the ion PF for the
$6$th (blue curves) and $10$th (green curves) ions without considering
background heating. In (a), two solid lines represent coarse-grained
values since the actual profiles contain fast oscillations as seen
in the two insets at $t=10^{4}t_{0}$ (left) and $t=10^{5}t_{0}$
(right), respectively. The coarse graining interval is $\Delta t/t_{0}=20$.
Note that in both insets the time span is $\Delta t^{\prime}/t_{0}=10$,
and the wavy (green) behavior belongs to the $10$th ion while the
PF of the $6$th ion stays nearly a constant (blue). To show the effect
of background heating, the PF profile of the $10$th ion is also plotted
for comparison (with $\kappa=10^{-4}$). That of the $6$th ion is
not explicitly shown because it is almost identical to the blue solid
(no-heating) curve. In (b), the fast oscillation amplitude is in the
order of $10^{-9}d_{0}$ so the dotted curves appear to be on top
of the solid ones. A thick and a thin (magenta) dotted lines representing
the $10$th and $6$th ions, respectively, with background heating
are also shown for comparison. (c)(d) The snapshots of the time-averaged
distributions. The relevant parameters are the same as Fig. \protect\ref{fig:diff_Tbg}.}
\label{fig:tev} 
\end{figure}

So far we have only discussed the steady-state solution to Eq. (\ref{eq:Variance}).
In this section we discuss the relaxation time scale towards the steady
state, which is also an important factor concerning the feasibility
of employing the sympathetic cooling in experiments. To illustrate
the general feature, we first calculate the dynamics of an $N=20$
ion chain in a harmonic trap under the edge cooling. We assume Doppler
cooling is applied to $5$ ions on each end of the chain and the whole
chain is initially in thermal equilibrium with temperature $T=2T_{D}$.
We then plot the curves of $\delta x_{i}$ and $\delta z_{i}$ with
$i=6$ (right next to the cooling ions) and $i=10$ (the middle ion)
as indicators in Fig. \ref{fig:tev}(a) and (b). Note that for the
axial motion the two solid lines have been coarse grained by a small
time interval. This is because the actual profiles have very fast
oscillations (see the insets of Fig. \ref{fig:tev}(a)). We also show
the upper and lower envelopes of such oscillations by the dotted lines.
The coarse-grained curves asymptotically approach constant values
as time increases, along with the fast oscillations dying away gradually.
We define a relaxation time $\tau_{R}$, beyond which the upper envelope
falls within $1\%$ of the coarse-grained value. So we find $\tau_{R}/t_{0}\sim10^{5}$
for the system to approach the steady state, where $t_{0}\equiv2\pi/\omega_{0}$
($\sim7\mu$s for most of the cases discussed here). For the transverse
direction, the amplitude of the fast oscillation is small, but it
takes $\tau_{R}/t_{0}\sim10^{6}$ to reach the steady state.

Now we consider the case with background heating at a rate $\kappa=\gamma_{bg}T_{bg}=10^{-4}$.
Different curves corresponding to this case are plotted in dashed
lines. Different $\gamma_{bg}$ change the final distribution of $\delta x_{i}$
and $\delta z_{i}$, but do not lead to significant variation of the
relaxation time scale. Fig. \ref{fig:tev}(c) and (d) shows the snapshots
of the distribution of $\delta x_{i}$ and $\delta z_{i}$ (coarse
graining also applied to the axial mode) at different times. The cooling
ions immediately reach their steady states (in a short time scale
$\gamma^{-1}$ which is not visible from the curve). The cooling then
starts to propagate to the inner part of the ion chain.

\begin{figure}
\begin{centering}
\includegraphics[width=8.5cm]{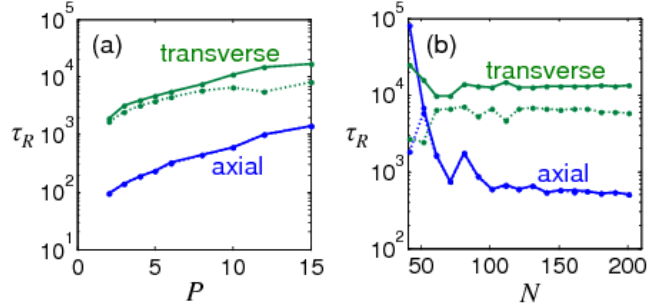} 
\par\end{centering}

\protect\caption{(Color online) The relaxation time $\tau_{R}$ as a function of (a)
the period $P$ in the periodic-node cooling for $N=121$, and (b)
the total length $N$ of the ion chain for a given $P=10$. The solid
curves correspond to no background heating cases and the dotted lines
correspond to background heating cases with $\kappa=10^{-4}$. For
(a), the relevant parameters are the same as Fig. \protect\ref{fig:periodiccool}.
For chains of different sizes in (b), the quartic trap is determined
by minimizing $\sum_{i=16}^{N-16}(z_{i+1}^{0}-z_{i}^{0})^{2}/(N-31)$
and setting $\sum_{i=16}^{N-16}(z_{i+1}^{0}-z_{i}^{0})/(N-31)=d_{0}$.}
\label{fig:scaling} 
\end{figure}

Previous discussion has shown that the relaxation time of edge cooling
is still quite long (0.1 to 1 second). We then turn to the more efficient
periodic-node cooling for a large ion chain. We here consider a quartic
trap and examine the relaxation time $\tau_{R}$ as a function of
the period $P$. As expected, Fig. \ref{fig:scaling}(a) shows that
$\tau_{R}$ is in general an increasing function with $P$. For an
$N=121$ chain, we find the time scale can be controlled within $\tau_{R}/t_{0}\sim10^{4}$
(tens of milliseconds) while the axial relaxation takes roughly $10$
times shorter than the transverse one. These results show a timescale
comparable to usual Doppler cooling (of order of a few milliseconds).
If the background heating is included, the transverse curve drops
slightly but the axial curve is hardly affected. As more ions are
added into the system, it is important to make sure that the relaxation
time does not scale up too fast with $N$. We show in Fig. \ref{fig:scaling}(b)
the scaling curves of $\tau_{R}$ with increasing $N$ by fixing $P=10$.
The axial relaxation time tends to decrease as the system size increases
and meets a lower bound in the large $N$ limit. On the contrary,
the transverse relaxation time appears to be independent of $N$.
This is because the transverse motion tends to involve only nearby
ions. So a longer chain is nothing but a simple repetition of segments
of a size $P$.

\section{Conclusion\label{sec:Conclusion}}

To conclude, we have presented a detailed investigation on the sympathetic
cooling in a large ion chain. Many findings discovered in this paper
are instructional for experimental implementation. First, a steady
state can be reached for a system subject to constant background heating
under continuous sympathetic cooling. By arranging cooling ancillary
ions in different ways, the cooling performance can be improved and
optimized. In our calculation, although the transverse motion is relatively
harder to be cooled than the axial one, by inserting ancillary ions
evenly over the chain it can be cooled down to a satisfactory level.
We have studied the effect of cooling rates and found the optimal
window of the cooling rates. We have also discussed the relaxation
dynamics and showed that the required time scale is within the reach
of experiments.

\bibliographystyle{apsrev}
\bibliography{symcool}

\end{document}